\documentstyle[11pt]{article}
\begin{document}
\begin{large}
\begin{titlepage}

\vspace{0.2cm}

\title{Neutralino pair production at proton-proton collider
\footnote{The project supported by National Natural Science
          Foundation of China}}
\author{{ Han Liang$^{a,b}$, Ma Wen-Gan$^{a,b}$, Jiang Yi$^{a}$,
          Zhou Mian-Lai$^{a}$ and Zhou Hong$^{a}$}\\
{\small $^{a}$Department of Modern Physics, University of Science
        and Technology}\\
{\small of China (USTC), Hefei, Anhui 230027, China.}\\
{\small $^{b}$Institute of Theoretical Physics, Academia Sinica,} \\
{\small P.O.Box 2735, Beijing 100080, China.} }
\date{}
\maketitle

\vskip 12mm

\begin{center}\begin{minipage}{5in}

\vskip 5mm
\begin{center} {\bf Abstract}\end{center}
\baselineskip 0.3in
{
In this paper we investigated the Drell-Yan process of the
light neutralino pair $\tilde{\chi}^0_{i}\tilde{\chi}^0_{j}~(i,j=1,2)$
productions at hadron colliders. We studied the dependence of the coupling
properties of two light neutralino $\tilde{\chi}^0_{1,2}$ on the three SUSY
Lagrangian parameters $M_1$, $M_2$ and $\mu$,
and find that the production rate of $\tilde{\chi}^0_1 \tilde{\chi}^0_2$
pair will be dominated by the Higgsino-like couplings under the condition
$\mu \ll M_2$ and $\mu \sim M_1$, while the productions of
$\tilde{\chi}^0_1\tilde{\chi}^0_1$ and $\tilde{\chi}^0_2\tilde{\chi}^0_2$
pairs will be enhanced by the gaugino-like couplings under $M_i \ll \mu$.
For the Higgsino-like $\tilde{\chi}^0_{1,2}$ neutralinos, the cross section
of $\tilde{\chi}^0_1\tilde{\chi}^0_2$ production at the LHC can reach $100~fb$.
}\\
\vskip 5mm
{PACS number(s): 14.80.Ly, 12.60.Jv}
\end{minipage}
\end{center}
\end{titlepage}

\baselineskip=0.36in

\eject
\rm
\baselineskip=0.36in

\begin{flushleft} {\bf 1. Introduction} \end{flushleft}
\par
With various theoretically appealings, the Minimal Supersymmetric
Standard Model (MSSM) is considered as the most potential choice for new
physics beyond the Standard Model(SM). However, due to the absence of
supersymmetric particle discovery in the energy range of current colliders,
the supersymmetry (SUSY) is clearly not an exact symmetry of nature, and
therefore must be broken. If R-parity conservation is hold and the SUSY
breaking is around TeV scale, the first two neutralinos $\tilde{\chi}^0_{1,2}$
are expected to be light, and the lightest one $\tilde{\chi}_1^0$ is probably
the lightest supersymmetric particle(LSP) in the SUSY models. Neutralinos
would be the most promising particles for the first experimental SUSY test.
Due to the high luminosity and large energy, the Large Hadron
Collider(LHC) has the ability to produce neutralino pairs
$\tilde{\chi}^0_i \tilde{\chi}^0_j$. In proton-proton collisions,
$\tilde{\chi}^0_i \tilde{\chi}^0_j$ pairs can be produced via Drell-Yan
subprocess and gluon-gluon fusion. Although the anti-quark luminosity
in distribution function of proton is much lower than gluon, yet the cross
sections of the neutralino pair productions via Drell-Yan mechanism are
competitive with those from the gluon-gluon fusion, since the former mechanism
of neutralino pair productions are accessible at the tree level. These facts
make the production rates in Drell-Yan process are competitive with or even
larger than those in gluon-gluon fusions. Therefore, it is significant in
theoretical and experimental studies of $\tilde{\chi}^0_i \tilde{\chi}^0_j$
pair productions via $q\bar{q}$ collisions at LHC.
\par
There are four SUSY model parameters involved in chargino/neutralino(ino)
sector, i.e. gaugino mass parameter $M_1$ and $M_2$, Higgs
mass parameter $\mu$ and the ratio of vacuum expectation values(VEV)
$\tan{\beta}=v_2/v_1$. These four parameters are essential ingredients of
the model: the two gaugino mass terms quantify the supersymmetry soft
breaking of $SU(2) \times U(1)$ subgroup, and together with $\mu$ and
$\tan{\beta}$ determine all the phenomenology properties in the ino sector.
In order to extract the ino physical quantities (i.e. masses, mixings and
couplings) from experimental observables (such as cross sections, branching
ratios and L-R asymmetries), one needs to study thoroughly the relations of
these quantities and observables with the fundamental SUSY parameters.
The discussion of the pair productions of neutral
gaugino and Higgsino current eigenstates can be found in Ref.\cite{photino}.
However, the discussions were not based on physical neutralino particles,
and thus could not really provide the sufficient information for the
investigation of SUSY particles.
In this work, we studied three typical
pair production processes of neutralino mass eigenstates in proton-proton
collisions, and investigated the dependence of their
cross sections on SUSY Lagrangian parameters. The paper is organized as
follows: in the section below, we present some general discussions on
the diagonalization of ino mass matrixes.
Sec. III gives the analytical tree-level formula of the cross
sections of subprocesses $q\bar{q} \rightarrow \tilde{\chi}^0_i
\tilde{\chi}^0_j$. In Sec.IV, we study the production rates of
$\tilde{\chi}^0_i\tilde{\chi}^0_j$ pairs in both
the subprocesses and the parent processes, and discuss the
dependence of the production rates on the SUSY model parameters.
Finally, a brief summary is presented.

\begin{flushleft} {\bf 2. The MSSM parameters in
                          chargino/neutralino sector} \end{flushleft}
\par
The charginos $\tilde{\chi}_j^+~(j=1,2)$ mass matrix in the current eigenstate
basis has the form
$$
X = \left(
      \begin{array}{cc}
         M_2 & \sqrt{2}\sin\beta m_W  \\
         \sqrt{2}\cos\beta m_W &  |\mu|e^{i\phi_{\mu}}
      \end{array}
      \right)
\eqno{(2.1a)}
$$
and the neutralino $\tilde{\chi}^0_i~(i=1,4)$ mass matrix is
$$
      Y = \left( \begin{array}{cccc}
     M_1 e^{i\phi_1} & 0 & -m_Z\sin\theta_{W}\cos\beta
                        & m_Z\sin\theta_{W}\sin\beta  \\
     0 & M_2 & m_Z\cos\theta_{W}\cos\beta
                        & -m_Z\cos\theta_{W}\sin\beta  \\
    -m_Z\sin\theta_{W}\cos\beta & m_Z\cos\theta_{W}\cos\beta & 0 & |\mu| e^{i\phi_{\mu}} \\
     m_Z\sin\theta_{W}\sin\beta & -m_Z\cos\theta_{W}\sin\beta & |\mu| e^{i\phi_{\mu}} & 0
      \end{array} \right).
\eqno{(2.1b)}
$$
Without loss of any generality, $\mu$ and $M_1$ can be complex which can
introduce CP-violation, and $M_2$ would be set to real and positive for its
phase angle can be rotated away by the field redefinition and thus absorbed
into $\phi_{\mu}$ and $\phi_{1}$. The diagonalizations of matrix $X$ and
$Y$, which give all the ino masses and their couplings, can be achieved with
$$
      U^{\ast}XV^{\dag} = X_{D}
$$
$$
      N^{\ast}YN^{\dag} = Y_{D}
\eqno{(2.2)}
$$
\par
In this work, we consider the ino sector with the following assumptions:
\begin{itemize}
\itemsep 0pt
\leftmargin .3cm
\item
{For simplification, CP-conservation is hold, namely $\phi_{\mu}=\phi_1=0$.
}

\item
{
The physical signs among $M_1$, $M_2$ and $\mu$ are relative, which
can be absorbed into phases $\phi_{\mu}$ and $\phi_1$ by redefinition of
fields. Thus, $M_1$, $M_2$ and $\mu$ are chosen to be real and positive,
i.e., $M_1, M_2, \mu > 0$.
}

\item
{
$\tan\beta$ is always set as an input parameter,
supposed that adequate information about $\tan{\beta}$ could be
obtained from some other experiments beyond ino sector.
}
\end{itemize}
With the above assumptions, generally we have two ways to choose
the input parameters in our calculation. One can employ the
scenario of taking $M_1$, $M_2$, $\mu$ and $\tan\beta$ as input parameters,
and then get all the physical ino masses and the matrix elements of $U$, $V$
and $N$ as outputs. The detail analyses can be found in Ref.\cite{Egypt}
\cite{zmlUVN1}\cite{Zerwas}. Since the Lagrangian parameters should be
extracted from the physical quantities, one can also choose an alternative
way to diagonalize the mass matrix $Y$ by taking any three physical ino
masses together with $\tan\beta$ as inputs. This strategy will provide
the other three ino masses and all the mixings and couplings as outputs.
There are several scenarios about the choices of three ino masses\cite{french}.
On the consideration that adequate information from chargino sector will give
out the two mass values of $\tilde{\chi}^+_{1,2}$, and from the energy
distribution of the final particles in the decay of $\tilde{\chi}^+_{1,2}$
at least one mass value of $m_{\tilde{\chi}^0_{i}}$ can be measured,
in this paper we take two chargino masses $m_{\tilde{\chi}^{+}_{1,2}}$ and
one of the neutralino masses $m_{\tilde{\chi}^{0}_{i}}$ as inputs.
In this way, the two fundamental SUSY parameter $M_2$ and $\mu$
can be figured out from the chargino masses by using following formula
$$
(M_2, \mu)=M_{\pm}=\frac{1}{2} \left(
        \sqrt{(m_{\tilde{\chi}^{+}_{1}}+m_{\tilde{\chi}^{+}_{2}})^2 -
              2 m_W^2 (1 - \sin{2\beta})}
    \pm \sqrt{(m_{\tilde{\chi}^{+}_{1}}-m_{\tilde{\chi}^{+}_{2}})^2 -
              2 m_W^2 (1 + \sin{2\beta})} \right).
\eqno{(2.3)}
$$
The ambiguity in determining $\mu$ and $M_{2}$ from the two-fold values of
$M_{\pm}$ can be cleaned up with some favourable measurements on the
chargino phenomenology. Practically, if the behaviours of charginos and
their couplings are Higgsino-like, there exists $M_2 \gg \mu$, while if
they are gaugino-like, we have $M_2 \ll \mu$. As to $M_1$, it is usually
a free parameter in the MSSM, if we have no further assumption. In the
second input strategy, $M_1$ can be determined by one of neutralino masses
$m_{\tilde{\chi}^{0}_{i}}$ alone, when $M_2$ and $\mu$ are given by Eq.(2.3).
\begin{eqnarray*}
&M_1& = \frac{1}{2} \left\{\mp 2 m_{\tilde{\chi}^{0}_{i}}
        (m_{\tilde{\chi}^{0}_{i}}^2 - \mu^2) (m_{\tilde{\chi}^{0}_{i}} \mp M_2)
        ~\pm~ 2 m_{\tilde{\chi}^{0}_{i}}^2 m_Z^2 ~-~
        2 m_{\tilde{\chi}^{0}_{i}} M_2 m_Z^2 \sin^2\theta_W ~+~ \right. \\
&~& \hskip2cm \left. \mu (2 m_{\tilde{\chi}^{0}_{i}} \mp M_2) m_Z^2 \sin2\beta ~\pm~
        \mu M_2 m_Z^2 \sin2\beta \cos2\theta_W \right\} / \\
&& \left\{ (\mu^2 - m_{\tilde{\chi}^{0}_{i}}^2)
           (m_{\tilde{\chi}^{0}_{i}}\mp M_2) ~+~
m_Z^2 \cos^2\theta_W (m_{\tilde{\chi}^{0}_{i}} \pm \mu \sin 2 \beta) \right\}
\end{eqnarray*}
$$
\eqno{(2.4)}
$$
Generally there are two solutions for $M_1$, and one is positive while the
other is negative. With the assumptions mentioned above, we take the positive
value solution from Eq.(2.4), and fix $M_1$ definitely.
Therefore, from the input mass values of two charginos ($m_{\chi^{+}_{1,2}}$)
and one of the neutralino mass($m_{\chi^0_i}$), one can extract all the SUSY
mass parameters $M_1$ and $M_{\pm}$
which denote $M_2$ and $\mu$ alternatively,
and then figure out the mass spectra and all the couplings of neutralino
sector consequently. In the following calculation and discussion, we will
adopt both input strategies discussed above.

\begin{flushleft} {\bf 3. Neutralino pair productions in $q\bar{q}$ collisions}
\end{flushleft}
\par
The Neutralino pair productions processes via the collisions of quark and
anti-quark in protons, can be expressed as
$$
q(p_1) \bar{q}(p_2) \rightarrow \tilde{\chi}^0_i(k_1) \tilde{\chi}^0_j(k_2)
~~~~~ (i,j=1,2,3,4)
\eqno{(3.2)}
$$
where $p_1$ and $p_2$ represent the momenta of the incoming quark and
anti-quark, and $k_1$ and $k_2$ denote the momenta of the two final state
neutralinos, respectively. The Mandelstam variables $\hat{s}$, $\hat{t}$ and
$\hat{u}$ are defined as $\hat{s}=(k_{1}+k_{2})^2$, $\hat{t}=(p_1-k_1)^2$,
$\hat{u}=(p_1-k_2)^2$. The relevant Feynmann diagrams are drawn in Fig.1.
The interaction Lagrangians involved are listed as \cite{HaberNucl}
\cite{Poland}
$$
{\cal L}_{Zq\bar{q}} = \frac{g}{\cos \theta_W} \cdot \bar{q} \gamma^{\mu}
                       [a_{Zq}^S + a_{Zq}^L P_L] q Z_{\mu}
$$
$$
{\cal L}_{Z\chi^0\chi^0} = \frac{g}{2 \cos \theta_W} \cdot \bar{\chi}^0_i
                            \gamma^{\mu} [O_Z^{ij} P_L - O_Z^{ij~\star} P_R]
                            \chi^0_j Z_{\mu}
$$
$$
{\cal L}_{q\tilde{q}\chi^0} = -\sqrt{2} g \cdot \bar{q}
                            [a^L_{qki} P_L + a^R_{qki} P_R]
                            \chi^0_i \tilde{q}_k
\eqno{(3.1a)}
$$
with the coupling constants
$$
a_{Zq}^S= Q_q \sin^2 \theta_W,~~~~
a_{Zq}^L= \frac{1}{2} (-1)^{T^3_q+1/2}
$$
$$
O_Z^{ij} = N_{i,4} N_{j,4}^{\star} - N_{i,3} N_{j,3}^{\star}
$$
$$
a^L_{qki} = \frac{m_q N_{i,5-q}^{\star}}{2 m_W \beta_q} R^{\star}_{q~k,1} -
        \tan \theta_W Q_q N_{i,1}^{\star} R^{\star}_{q~k,2}
$$
$$
a^R_{qki} = (T^3_q N_{i,2} - \tan\theta_W (T^3_q-Q_q) N_{i,1}) R^{\star}_{q~k,1} +
        \frac{m_q N_{i,5-q}}{2 m_W \beta_q} R^{\star}_{q~k,2}
\eqno{(3.1b)}
$$
Here $P_{R,L}=(1\pm \gamma_5)/2$. $R_q$ denote the squark transformation
matrix. $q=1,2$ denote up-type and down-type quarks respectively, and
$$
\beta_q =
\left\{
\begin{array}{cc}
\sin \beta & q=1 \\
\cos \beta & q=2
\end{array}
\right.
\eqno{(3.1c)}
$$
The corresponding Lorentz invariant matrix element for the tree-level
process is written as
$$
{\cal M}_{0}= {\cal M}_{\hat{s}}+ {\cal M}_{\hat{t}}+ {\cal M}_{\hat{u}}
\eqno{(3.3a)}
$$
where
$$
{\cal M}_{\hat{s}} = \frac{i g^2}{2 \cos^2 \theta_W \cdot (\hat{s} - m_Z^2)}
        \bar{u}(k_1) \gamma^{\mu} [O_Z^{ij} P_L - O_Z^{ij\star} P_R] v(k_2) \cdot
        \bar{v}(p_2) \gamma_{\mu} [a_{Zq}^{S} + a_{Zq}^{L} P_L] u(p_1)
$$
$$
{\cal M}_{\hat{t}} = \frac{-2 i g^2}{\hat{t} - m_{\tilde{q}_k}^2}
        \bar{u}(k_1) (a^{L\star}_{qki} P_R + a^{R\star}_{qki} P_L) u(p_1) \cdot
        \bar{v}(p_2) (a^{L}_{qkj} P_L + a^{R}_{qkj} P_R) v(k_2)~~~(k=1,2)
$$
$$
{\cal M}_{\hat{u}} = (-1)^{\delta_{ij}} \frac{-2 i g^2}{\hat{u} - m_{\tilde{q}_l}^2}
        \bar{u}(k_2) (a^{L\star}_{qlj} P_R + a^{R\star}_{qlj} P_L) u(p_1) \cdot
        \bar{v}(p_2) (a^{L}_{qli} P_L + a^{R}_{qli} P_R) v(k_1)~~~(l=1,2)
\eqno{(3.3b)}
$$
Here we take simply vanished masses of the first generation quarks, i.e.
$m_u=m_d=0$. The relative sign $(-1)^{\delta_{ij}}$ of ${\cal M}_{\hat{u}}$
towards ${\cal M}_{\hat{t}}$ and ${\cal M}_{\hat{s}}$ is merely due to Pauli
statistics.
\par
The corresponding differential cross section at tree level can be written as
$$
\frac{d \sigma}{d \Omega} = \frac{1}{4}\frac{N_c}{9}(\frac{1}{2})^{\delta_{ij}}
        \frac{g^4 \lambda_{ij}}{32 \pi^2 \hat{s}^2}
        \left( I_{\hat{s}\hat{s}} + I_{\hat{t}\hat{t}} + I_{\hat{u}\hat{u}} +
               2 I_{\hat{s}\hat{t}} +
               2 (-1)^{\delta_{ij}} I_{\hat{s}\hat{u}} +
               2 (-1)^{\delta_{ij}} I_{\hat{t}\hat{u}} \right)
\eqno{(3.4a)}
$$
where the factors $\frac{1}{4}$, $\frac{N_c}{9}$ and
$(\frac{1}{2})^{\delta_{ij}}$ are the initial spin-average, color-average and
final identical-particle factors respectively. The squares of matrix element
have the form as
\begin{eqnarray*}
I_{\hat{s}\hat{s}} &=& \frac{2}{\cos^4 \theta_W (\hat{s}-m_Z^2)^2}
        (a_{Zq}^{S~2} + a_{Zq}^S a_{Zq}^L + \frac{1}{2} a_{Zq}^{L~2}) \\
        &~& \left[O_Z^{ij} O_Z^{ij\star} \cdot ((m_{\tilde{\chi}_i}^2 - \hat{t})(m_{\tilde{\chi}_j}^2 - \hat{t}) +
                   (m_{\tilde{\chi}_i}^2 - \hat{u})(m_{\tilde{\chi}_j}^2 - \hat{u})) -
         (O_Z^{ij~2} + O_Z^{ij\star~2}) \cdot m_{\tilde{\chi}_i} m_{\tilde{\chi}_j} \hat{s} \right]
\end{eqnarray*}
\begin{eqnarray*}
I_{\hat{t}\hat{t}} &=& \frac{4}{(\hat{t}-m_{\tilde{q}_k}^2)
                                (\hat{t}-m_{\tilde{q}_{k^{\prime}}}^2)}
        (a^L_{qk^{\prime}i} a^{L\star}_{qki} + a^R_{qk^{\prime}i} a^{R\star}_{qki})
        (a^L_{qkj} a^{L\star}_{qk^{\prime}j} + a^R_{qkj} a^{R\star}_{qk^{\prime}j}) \cdot
        (m_{\tilde{\chi}_i}^2 - \hat{t}) (m_{\tilde{\chi}_j}^2 - \hat{t})
\end{eqnarray*}
\begin{eqnarray*}
I_{\hat{u}\hat{u}} &=& \frac{4}{(\hat{u}-m_{\tilde{q}_l}^2)
                                (\hat{u}-m_{\tilde{q}_{l^{\prime}}}^2)}
        (a^L_{ql^{\prime}j} a^{L\star}_{qlj} + a^R_{ql^{\prime}j} a^{R\star}_{qlj})
        (a^L_{qli} a^{L\star}_{ql^{\prime}i} + a^R_{qli} a^{R\star}_{ql^{\prime}i}) \cdot
        (m_{\tilde{\chi}_i}^2 - \hat{u}) (m_{\tilde{\chi}_j}^2 - \hat{u})
\end{eqnarray*}

\begin{eqnarray*}
I_{\hat{s}\hat{t}} &=& \frac{-2}{\cos^2 \theta_W (\hat{s}-m_Z^2) (\hat{t}-m_{\tilde{q}_k}^2)} \\
        &~& \left[ (a_{Zq}^{S} a^L_{qki} a^{L\star}_{qkj} O_Z^{ij} -
              (a_{Zq}^S+a_{Zq}^L) a^R_{qki} a^{R\star}_{qkj} O_Z^{ij\star}) \cdot
           (m_{\tilde{\chi}_i}^2 - \hat{t}) (m_{\tilde{\chi}_j}^2 - \hat{t}) + \right. \\
     &~& \left. ((a_{Zq}^S+a_{Zq}^L) a^R_{qki} a^{R\star}_{qkj} O_Z^{ij} -
              a_{Zq}^S a^L_{qki} a^{L\star}_{qkj} O_Z^{ij\star}) \cdot
           m_{\tilde{\chi}_i} m_{\tilde{\chi}_j} \hat{s} \right]
\end{eqnarray*}
\begin{eqnarray*}
I_{\hat{s}\hat{u}} &=& \frac{-2}{\cos^2 \theta_W (\hat{s}-m_Z^2) (\hat{u}-m_{\tilde{q}_l}^2)} \\
        &~& \left[ ((a_{Zq}^S+a_{Zq}^L) a^R_{qlj} a^{R\star}_{qli} O_Z^{ij} -
                    a_{Zq}^S a^L_{qlj} a^{L\star}_{qli} O_Z^{ij\star}) \cdot
                   (m_{\tilde{\chi}_i}^2 - \hat{u}) (m_{\tilde{\chi}_j}^2 - \hat{u}) + \right. \\
        &~& \left. (a_{Zq}^S a^L_{qlj} a^{L\star}_{qli} O_Z^{ij} -
                    (a_{Zq}^S+a_{Zq}^L) a^R_{qlj} a^{R\star}_{qli}
                    O_Z^{ij\star}) \cdot m_{\tilde{\chi}_i} m_{\tilde{\chi}_j} \hat{s} \right]
\end{eqnarray*}
\begin{eqnarray*}
I_{\hat{t}\hat{u}} &=& \frac{4}{(\hat{t}-m_{\tilde{q}_k}^2) (\hat{u}-m_{\tilde{q}_l}^2)} \\
        &~& \left[ (a^{L\star}_{qli} a^L_{qkj} a^{L\star}_{qki} a^L_{qlj} +
                    a^{R\star}_{qli} a^R_{qkj} a^{R\star}_{qki} a^R_{qlj}) \cdot
                    m_{\tilde{\chi}_i} m_{\tilde{\chi}_j} \hat{s} + \right. \\
        &~& \frac{1}{2} (a^{L\star}_{qli} a^L_{qkj} a^{R\star}_{qki} a^R_{qlj} +
                         a^{R\star}_{qli} a^R_{qkj} a^{L\star}_{qki} a^L_{qlj}) \cdot \\
        &~& \left. ((m_{\tilde{\chi}_i}^2 - \hat{t}) (m_{\tilde{\chi}_j}^2 - \hat{t}) +
                    (m_{\tilde{\chi}_i}^2 - \hat{u}) (m_{\tilde{\chi}_j}^2 - \hat{u}) -
                    (\hat{s} - m_{\tilde{\chi}_i}^2 - m_{\tilde{\chi}_j}^2) \hat{s}) \right]
\end{eqnarray*}
$$
\eqno{(3.4b)}
$$
where
$$
\lambda_{ij} = \sqrt{(\hat{s}-m_{\tilde{\chi}_i}^2-m_{\tilde{\chi}_j}^2)^2-4 m_{\tilde{\chi}_i}^2 m_{\tilde{\chi}_j}^2}/2
$$

\begin{flushleft} {\bf 4. Numerical results and discussion} \end{flushleft}
\par
As we assume $\tilde{\chi}^0_{1,2}$ are lighter than
$\tilde{\chi}^0_{3,4}$ and $\tilde{\chi}^0_1$ is likely to be the LSP,
the three types of channels: $q \bar{q} \rightarrow
\tilde{\chi}_1^0 \tilde{\chi}_1^0$,
$q \bar{q} \rightarrow \tilde{\chi}_2^0 \tilde{\chi}_2^0$ and
$q \bar{q} \rightarrow \tilde{\chi}_1^0 \tilde{\chi}_2^0$
(associated production) would be the most dominant neutralino pair production
processes which may lead to the first detection of SUSY particles at
the LHC\cite{HaberRep}. Here we present the numerical results of the cross
sections of these three processes and discuss their dependences on the
basic SUSY parameters. We divide the input MSSM parameters into two parts.
One is for for the general parameters included also in the SM,
and the other is the ino and squark sectors of the MSSM.
For the first parameter part, we take
$m_Z=91.1887~GeV$, $\sin^2 \theta_W=0.2315$, $\alpha=1/137.03598$.
For the second part, we just limit the values of $M_1$,
$M_2$ and $\mu$ to be real, positive and below $1~TeV$, and take
$\tan\beta=1.5$, $m_{\tilde{u}_1} = m_{\tilde{d}_1}=350~GeV$,
$m_{\tilde{u}_2} = m_{\tilde{d}_2} = 550~GeV$, and
$\theta_{\tilde{u}}= \theta_{\tilde{d}}=\pi/4$.
The three ino physical masses are taken as
$$
m_{\tilde{\chi}^{+}_{1}}=150~GeV,~~~
m_{\tilde{\chi}^{+}_{2}}=550~GeV,~~~
m_{\tilde{\chi}^{0}_{1}}=100~GeV,
\eqno{(4.1a)}
$$
\par
By using Eq.(2.3) with above ino mass values, one may have two choices
of parameter sets for $\mu$ and $M_{2}$, which are in two extreme cases
respectively, namely Higgsino-like and gaugino-like ino states.
For the Higgsino-like case, we get
$$
M_2=534~GeV,~~~ \mu=166~GeV,~~~ M_1=135~GeV,
$$
and other neutralino masses as
$$
m_{\tilde{\chi}^{0}_{2}}=166~GeV,~~~
m_{\tilde{\chi}^{0}_{3}}=184~GeV,~~~
m_{\tilde{\chi}^{0}_{4}}=550~GeV.
\eqno{(4.1b)}
$$
For the gaugino-like case, we have
$$
M_2=166~GeV,~~~ \mu=534~GeV,~~~ M_1=105~GeV,
$$
and other three neutralino masses as
$$
m_{\tilde{\chi}^{0}_{2}}=151~GeV,~~~
m_{\tilde{\chi}^{0}_{3}}=534~GeV,~~~
m_{\tilde{\chi}^{0}_{4}}=554~GeV.
\eqno{(4.1c)}
$$
For complete discussion, we present also the results for the mixture case.
We take the physical mass inputs as follows for the mixture ino states
$$
m_{\tilde{\chi}^{+}_{1}} = 122~GeV,~~~ m_{\tilde{\chi}^{+}_{2}} = 280~GeV,~~~
m_{\tilde{\chi}^{0}_{1}} = 101~GeV.
$$
the corresponding outputs obtained as
$$
M_2=\mu=200~GeV,~~~ M_1=150~GeV,
$$
$$
m_{\tilde{\chi}^{0}_{2}}=164~GeV,~~~ m_{\tilde{\chi}^{0}_{3}}=201~GeV,~~~
m_{\tilde{\chi}^{0}_{4}}=286~GeV,
\eqno{(4.1d)}
$$
Then the cross sections of the subprocesses $q\bar{q} \rightarrow
\tilde{\chi}^0_1 \tilde{\chi}^0_1, \tilde{\chi}^0_1 \tilde{\chi}^0_2,
\tilde{\chi}^0_2 \tilde{\chi}^0_2$ can be numerically evaluated.
The cross sections for the three subprocesses are plotted in Fig.2(a),(b)
and (c) respectively, as the functions of the parton-parton collision
c.m.s. energy $\sqrt{\hat{s}}$. In Fig.2(a) and (c), all the curves for
gaugino-like case show obviously the threshold effects when $\sqrt{\hat{s}}$
is just above the threshold energies of the $\tilde{\chi}^0_1\tilde{\chi}^0_1$
and $\tilde{\chi}^0_2\tilde{\chi}^0_2$ pair production, respectively.
The cross sections for Higgsino-like case are fairly smaller than the
corresponding ones for gaugino-like case in the process of
$\tilde{\chi}^0_i \tilde{\chi}^0_i$ pair production, and the curves
for the Higgsino-like $\tilde{\chi}^0_2 \tilde{\chi}^0_2$ pair production
are even too small to be plotted in Fig.2(c).
On the contrary, the Higgsino couplings will enhance
abruptly the cross sections for the processes of
$\tilde{\chi}^0_1 \tilde{\chi}^0_2$ pair production as shown in Fig.2(b),
but the cross sections for gaugino-like and mixture cases are negligibly
small compared with those for Higgsino-like case.
\par
The reactions of $q\bar{q} \rightarrow \tilde{\chi}^0_i \tilde{\chi}^0_j$
are only subprocesses of the parent $pp$ hadron collider. The total cross
sections of $\tilde{\chi}^0_i \tilde{\chi}^0_j$ pair productions via
$q\bar{q}$ annihilation in $pp$ collider can be simply obtained by folding
the cross sections of the subprocesses $\hat{\sigma} [q\bar{q} \rightarrow
\tilde{\chi}^0_i \tilde{\chi}^0_j]~~(q=u,d)$ with the quark and
anti-quark luminosity. Adopting quark and anti-quark structure
functions of the MRS set G given in Ref.\cite{MRSGFIT} and
$Q=\sqrt{\hat{s}}$, we calculate
the cross sections $\sigma[pp\rightarrow q\bar{q}\rightarrow
\tilde{\chi}^0_i\tilde{\chi}^0_j]~(i,j=1,2)$
as the functions of the $pp$ collider c.m.s energy $\sqrt{s}$,
under the same conditions of Eq.(4.1). The results are depicted in Fig.2(d).
It is impressive that the $\tilde{\chi}^0_1\tilde{\chi}^0_2$
pair production rate at the LHC, can even reach as large as
$1.5 \times 10^2 fb$, when the couplings are Higgsino-like.
We also calculate the $\tilde{\chi}^0_1\tilde{\chi}^0_2$
pair production at the Tevatron. The cross section $\sigma[p\bar{p}
\rightarrow q\bar{q} \rightarrow \tilde{\chi}^0_1 \tilde{\chi}^0_2]$
can be $0.02~pb$ in Higgsino-like case, when $\sqrt{s}=2~TeV$.
We see that the light neutralino pair production rates at the Tevatron
are far smaller than those at the LHC, due to the lower c.m.s energy at
Tevatron.
\par
In Fig.3(a),(b) and (c), the cross sections $\hat{\sigma}[u\bar{u} \rightarrow
\tilde{\chi}^0_1 \tilde{\chi}^0_1, \tilde{\chi}^0_1 \tilde{\chi}^0_2,
\tilde{\chi}^0_2 \tilde{\chi}^0_2]$ versus the parameter $M_2$, with
$\sqrt{\hat{s}}=1~TeV$ and $\mu=400~GeV$, are plotted by taking
$M_1=100,~400~GeV$ and $1~TeV$, respectively.
As discussed above with the physical masses $m_{\tilde{\chi}^{+}_{1,2}}$ and
one of $m_{\tilde{\chi}^{0}_{i}}$, one might get three model parameters $M_1$
and $M_{\pm}$ where $M_{\pm}$ denote $M_2$ and $\mu$ alternatively.
The properties of neutralinos depend not only on $M_2$
and $\mu$ as charginos do, but also on the mass parameter $M_1$.
From Fig.3, we can say in the case that when there is a large split between
$M_{\pm}$ $(M_+ \gg M_-)$, the $\tilde{\chi}^0_i\tilde{\chi}^0_j$ pair
productions via $q\bar{q}$ collisions have some interesting features
as follows:
\begin{itemize}
\itemsep 0pt
\leftmargin .3cm
\item
{ When $M_1 \gg M_{\pm}$, $\tilde{\chi}^0_{1,2}$ are mainly decide by $M_2$
and $\mu$:
if $\mu=M_- \ll M_2=M_+$, $\tilde{\chi}^0_{1,2}$ are dominantly composed of
Higgsino, and consequently $\tilde{\chi}^0_1\tilde{\chi}^0_2$ pair production
rate is significant while $\tilde{\chi}^0_i\tilde{\chi}^0_i~~(i=1,2)$ pair
productions are suppressed;
if $\mu=M_- \gg M_2=M_+$, $\tilde{\chi}^0_{1}$ is approximately $\tilde{W}_3$
and $\tilde{\chi}^0_{2}$ is Higgsino, and accordingly
$\tilde{\chi}^0_1\tilde{\chi}^0_1$ pair production is enhanced while
$\tilde{\chi}^0_1\tilde{\chi}^0_2$ and $\tilde{\chi}^0_2\tilde{\chi}^0_2$ pair
productions are suppressed.
Shown as in Fig.3 with the cases of $M_1=1~TeV$, and the curve of
$\tilde{\chi}^0_2\tilde{\chi}^0_2$ pair production for the case
$M_1=1~TeV$ is too small to be plotted in Fig.3(c).
}

\item
{
When $M_1 \ll M_{\pm}$, $\tilde{\chi}^0_{1}$ is $\tilde{B}$. Accordingly,
$\tilde{\chi}^0_1\tilde{\chi}^0_1$, $\tilde{\chi}^0_1\tilde{\chi}^0_2$
pair productions are dominated by gaugino-couplings, and the former is
enhance and the latter is suppressed (the exceptation for the curve of
$M_1=100~GeV$ in Fig.3(b) are merely because of the mass degeneration
between $m_{\tilde{\chi}^0_{1,2}}$ when $M_1=M_2=100~GeV$).
$\tilde{\chi}^0_{2}$ are decide by $M_2$ and $\mu$:
if $M_2=M_- \ll \mu=M_+$, $\tilde{\chi}^0_{2}$ is mainly $\tilde{W}_3$, and
$\tilde{\chi}^0_2\tilde{\chi}^0_2$ pair production is enhanced;
and vice versa.
Shown as in Fig.3 with the cases of $M_1=100~GeV$.
}

\item
{
When $M_- {}^{<}_{\sim} M_1 {}^{<}_{\sim} M_+$, the effects of $M_1$ on
$\tilde{\chi}^0_{1,2}$ are rather weak, and only lead to some mixture.
Then if $\mu=M_- \ll M_2=M_+$, $\tilde{\chi}^0_{1,2}$ are in some
Higgsino-like states, and consequently $\tilde{\chi}^0_1\tilde{\chi}^0_2$
pair production is significant while $\tilde{\chi}^0_i\tilde{\chi}^0_i$ (i=1,2)
pairs are suppressed;
if $M_2=M_- \ll \mu=M_+$, $\tilde{\chi}^0_{1}$ is in gaugino-like state
while $\tilde{\chi}^0_{2}$ may be in some mixture state, and accordingly
$\tilde{\chi}^0_1\tilde{\chi}^0_1$ pair production is
enhanced while $\tilde{\chi}^0_1\tilde{\chi}^0_2$,
$\tilde{\chi}^0_2\tilde{\chi}^0_2$ pairs are suppressed.
Shown as in Fig.3 with the cases of $M_1=400~GeV$.
}
\end{itemize}
These features can be concluded as that pure gaugino-couplings dominate
the production of $\tilde{\chi}^0_i\tilde{\chi}^0_i$ pair
(as shown by the curves in the areas $M_2 < \mu=400~GeV$ of Fig.3(a,c)),
while pure Higgsino couplings enhance the $\tilde{\chi}^0_i\tilde{\chi}^0_j~
(i\neq j)$ pair productions
(as shown by the curve of $M_1=1~TeV$ in the area $\mu \ll M_2$ of Fig.3(b)),
and any gaugino-like $\tilde{\chi}^{0}_{1}$ will spoil the large production
rate of $\tilde{\chi}^0_1\tilde{\chi}^0_2$.

\begin{flushleft} {\bf 5. Conclusions} \end{flushleft}
\par
In this work we studied the productions of the light neutralino pairs
$\tilde{\chi}^0_1 \tilde{\chi}^0_1$, $\tilde{\chi}^0_1 \tilde{\chi}^0_2$,
$\tilde{\chi}^0_2 \tilde{\chi}^0_2$ via Drell-Yan process
at hadron colliders, and investigated the correlations between the property
of the neutralino pair productions and the basic SUSY Lagrangian
parameter $M_1$, $M_2$ and $\mu$.
\par
From the numerical results, it can be concluded that Higgsino-couplings will
increase the $\tilde{\chi}^0_1\tilde{\chi}^0_2$ pair production significantly,
while gaugino-couplings enhance the $\tilde{\chi}^0_1\tilde{\chi}^0_1$
and $\tilde{\chi}^0_2\tilde{\chi}^0_2$ pair productions.
If $\mu << M_2$ and $\mu {}^{<}_{\sim} M_1$, Higgsino components will be
dominated in $\tilde{\chi}^0_1$ and $\tilde{\chi}^0_2$, and the cross section
of Higgsino-like $\tilde{\chi}^0_1\tilde{\chi}^0_2$ pair production at the
LHC, can reach $100~fb$. Thus, by taking a annual luminosity of $pp$
collision at the LHC being $100~fb^{-1}$, one can accumulate
$1 \times 10^4$ events per year. Therefore the precise measurement of this
process is suitable in detecting SUSY signals and helpful in determining the
basic SUSY parameters.

\vskip 4mm
\noindent{\large\bf Acknowledgement:}
These work was supported in part by the National Natural Science
Foundation of China(project numbers: 19675033, 19875049), the Youth Science
Foundation of the University of Science and Technology of China and
a grant from the State Commission of Science and Technology of China.

\vskip 5mm
\begin{flushleft} {\bf Figure Captions} \end{flushleft}

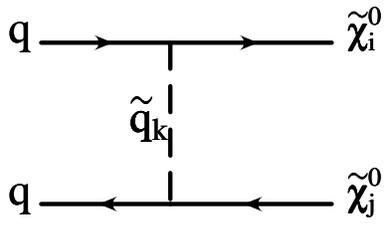
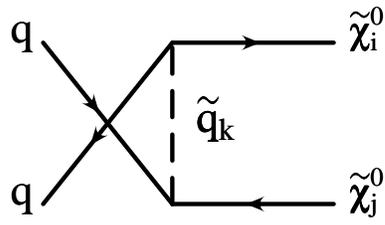
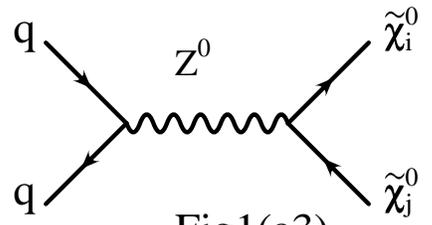
{\bf Fig.1} The Feynman diagrams of the subprocess $q\bar{q} \rightarrow
\tilde{\chi}^0_i\tilde{\chi}^0_j$.

{\bf Fig.2(a)} The cross sections of the subprocess $q\bar{q} \rightarrow
\tilde{\chi}^0_1\tilde{\chi}^0_1$ as functions of $\sqrt{\hat{s}}$.
The letter 'u' and 'd' denote the different processes with initial $u\bar{u}$
and $d\bar{d}$ collisions respectively, while 'H', 'g' and 'm' denote the
Higgsino-like, gaugino-like and mixture cases respectively.

{\bf Fig.2(b)} The cross sections of the subprocess $q\bar{q} \rightarrow
\tilde{\chi}^0_1\tilde{\chi}^0_2$ as functions of $\sqrt{\hat{s}}$.
The letter 'u' and 'd' denote the different processes with initial $u\bar{u}$
and $d\bar{d}$ collisions respectively, while 'H', 'g' and 'm' denote the
Higgsino-like, gaugino-like and mixture cases respectively.

{\bf Fig.2(c)} The cross sections of the subprocess $q\bar{q} \rightarrow
\tilde{\chi}^0_2\tilde{\chi}^0_2$ as functions of $\sqrt{\hat{s}}$.
The letter 'u' and 'd' denote the different processes with initial $u\bar{u}$
and $d\bar{d}$ collisions respectively, while 'H', 'g' and 'm' denote the
Higgsino-like, gaugino-like and mixture cases respectively.

{\bf Fig.2(d)} The total cross sections of the process $pp \rightarrow q\bar{q}
\rightarrow \tilde{\chi}^0_i\tilde{\chi}^0_j$ as functions of $\sqrt{s}$,
where 'H', 'g' and 'm' denote the Higgsino-like, gaugino-like and mixture
cases, respectively.

{\bf Fig.3(a)} The cross sections of the subprocess $u\bar{u} \rightarrow
\tilde{\chi}^0_1\tilde{\chi}^0_1$ as functions of $M_2$ with $\mu=400~GeV$
and $\sqrt{\hat{s}}=1~TeV$.

{\bf Fig.3(b)} The cross sections of the subprocess $u\bar{u} \rightarrow
\tilde{\chi}^0_1\tilde{\chi}^0_2$ as functions of $M_2$ with $\mu=400~GeV$
and $\sqrt{\hat{s}}=1~TeV$.

{\bf Fig.3(c)} The cross sections of the subprocess $u\bar{u} \rightarrow
\tilde{\chi}^0_2\tilde{\chi}^0_2$ as functions of $M_2$ with $\mu=400~GeV$
and $\sqrt{\hat{s}}=1~TeV$.

\end{large}

\vskip 5mm
\begin{thebibliography}{s30}
\bibitem{photino} J. Ellis and Graham G. Ross, Phys. Lett. B117(1982)397;
                  J.M. Frere and G.L. Kane, Nucl. Phys. B223(1983)331.
\bibitem{Egypt} Mohamed M.El Kheishen, Ahmed A. Shafik and Amr A. Aboshousha,
                Phys. Rev. D45(1992)4345-4345.
\bibitem{zmlUVN1} M.L. Zhou, W.G. Ma, L. Han, Y. Jiang and H. Zhou,
                 J. Phys. G25(1999)27-43
\bibitem{Zerwas} S.Y. Choi, A. Djouadi, H.S. Song and P.M. Zerwas, hep-ph/9812236.
\bibitem{french} J.L. Kneur and G. Moultaka, hep-ph/9807336.
\bibitem{HaberNucl} J.F. Gunion and H.E. Haber, Nucl. Phys. B272 (1986)1-76.
\bibitem{Poland} J. Rosiek, Phys. Rev. D41(1998)3464.
\bibitem{HaberRep} H.E. Haber and G.L. Kane, Phys. Rep. 117 (1985)75-263.
\bibitem{MRSGFIT} A.D. Martin, R.G. Roberts and W.J.Stirling,
                 Phys. Lett. B354(1995)155-162.
\end{thebibliography}
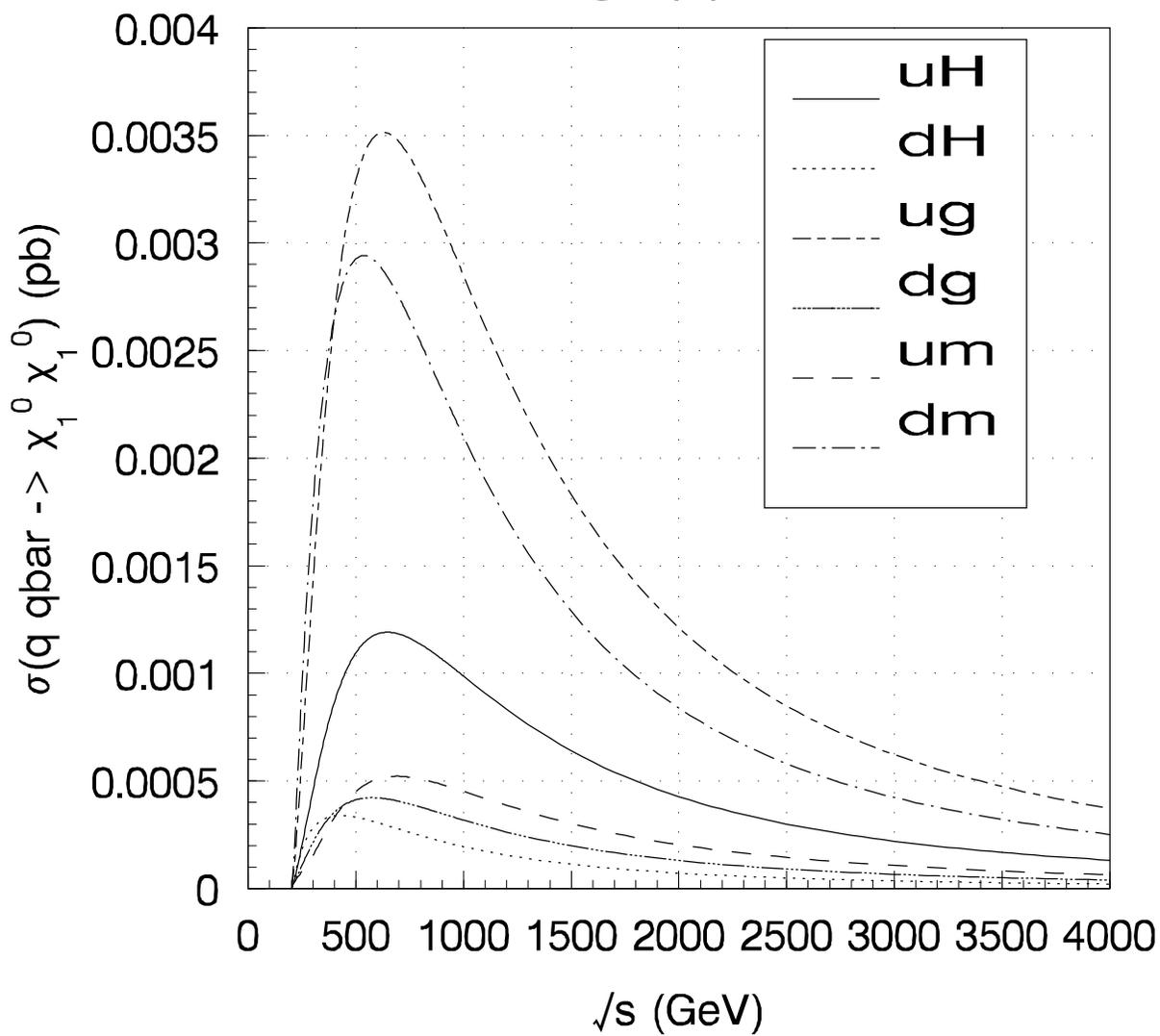
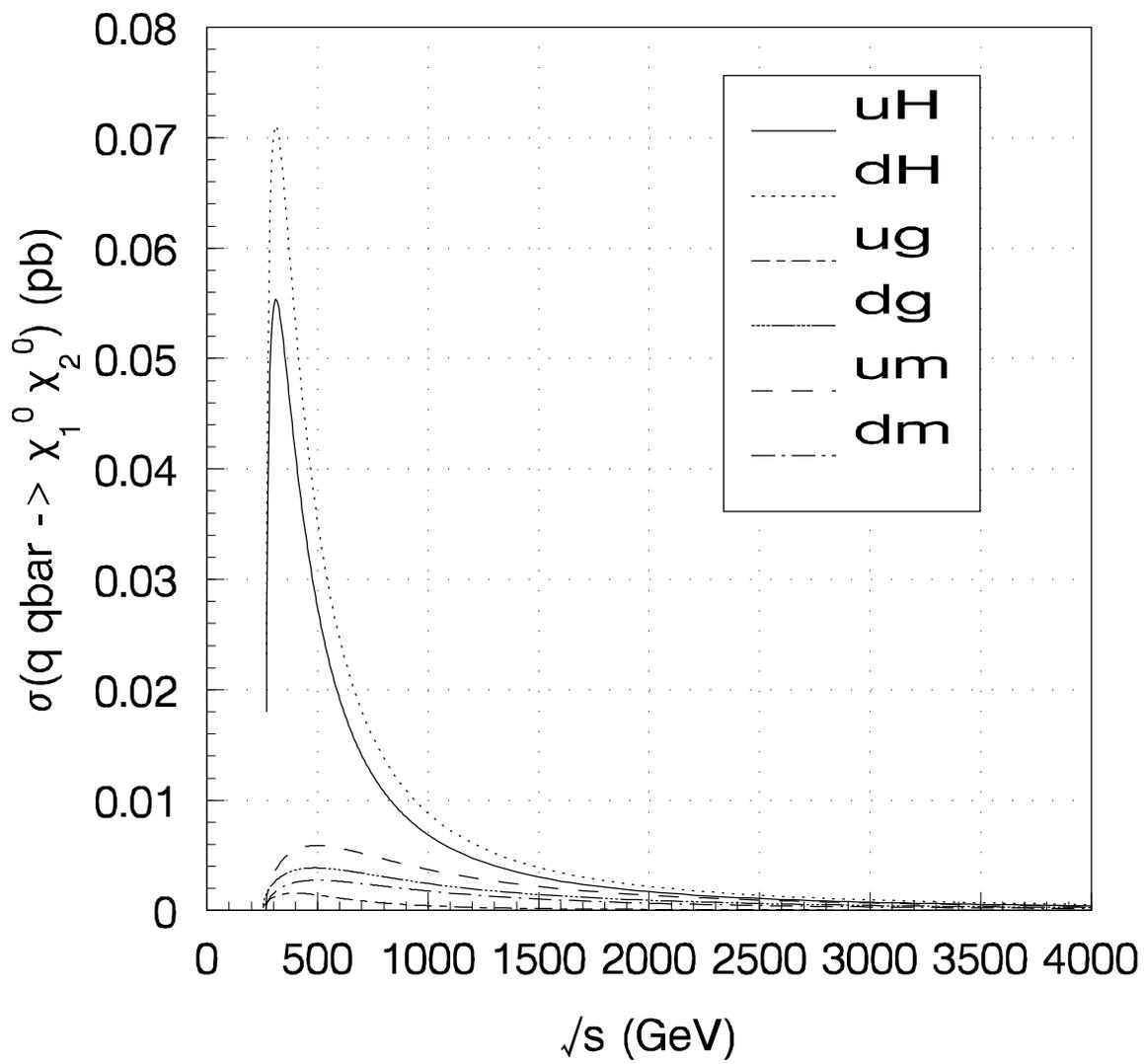
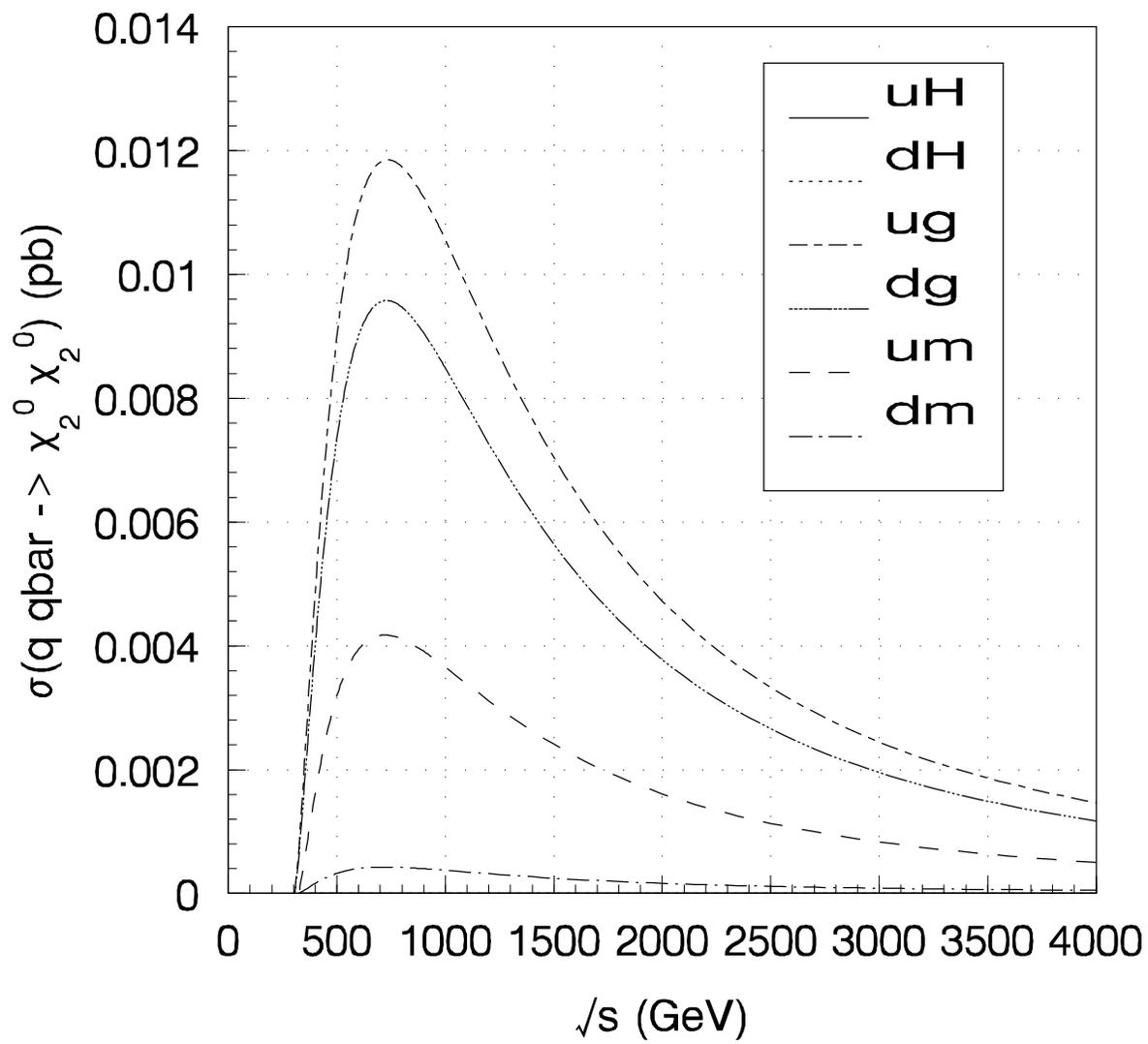
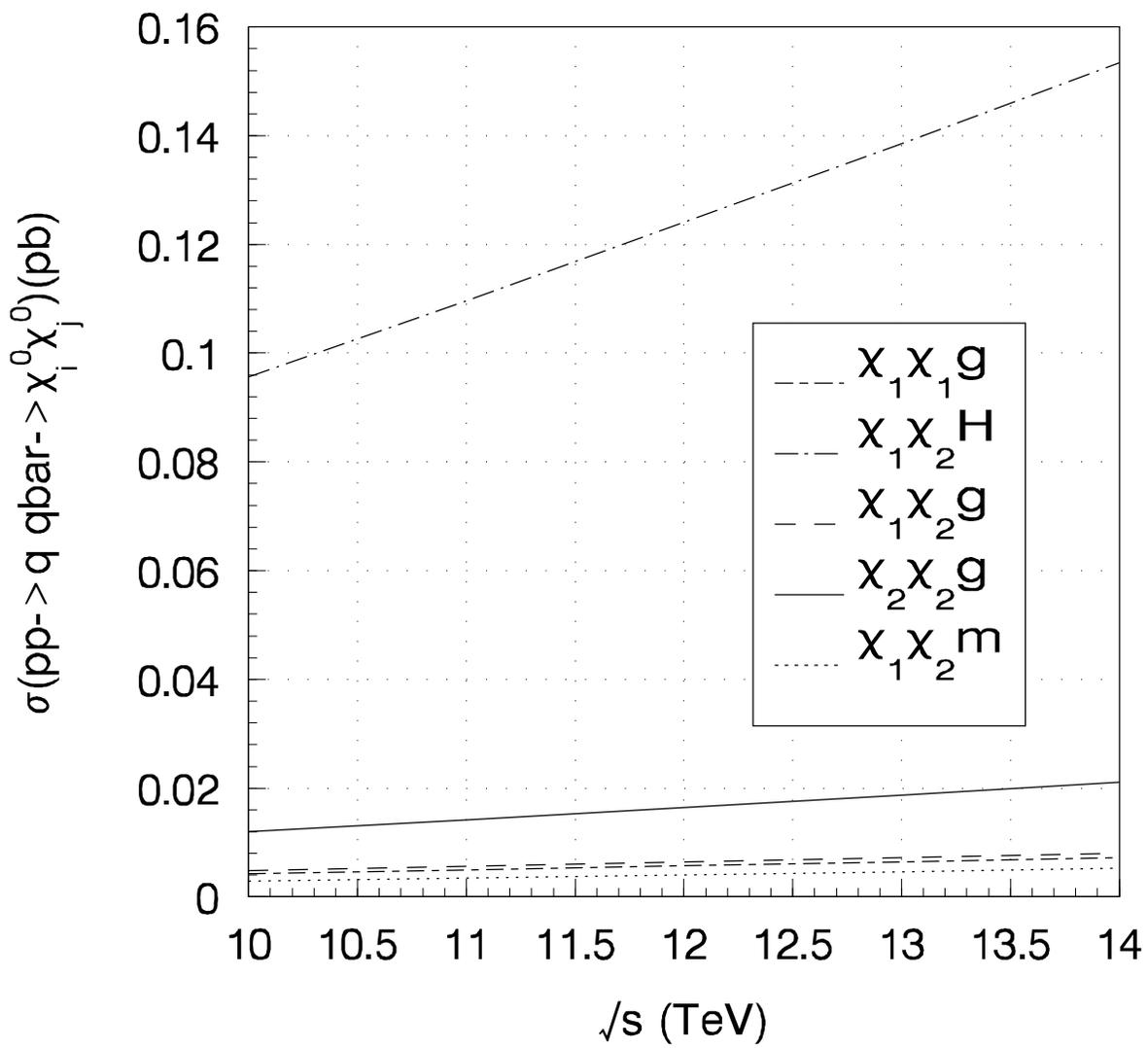
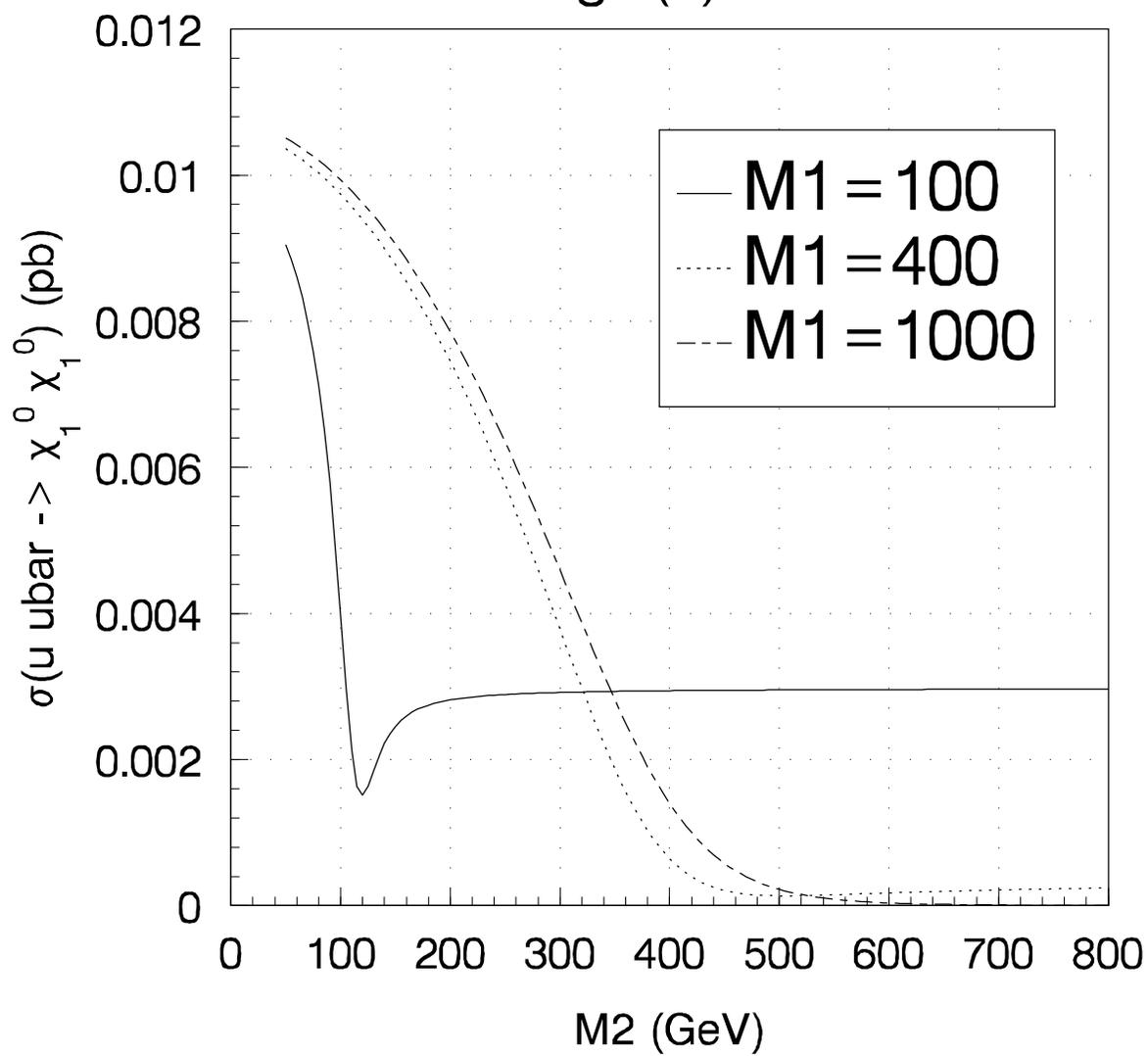
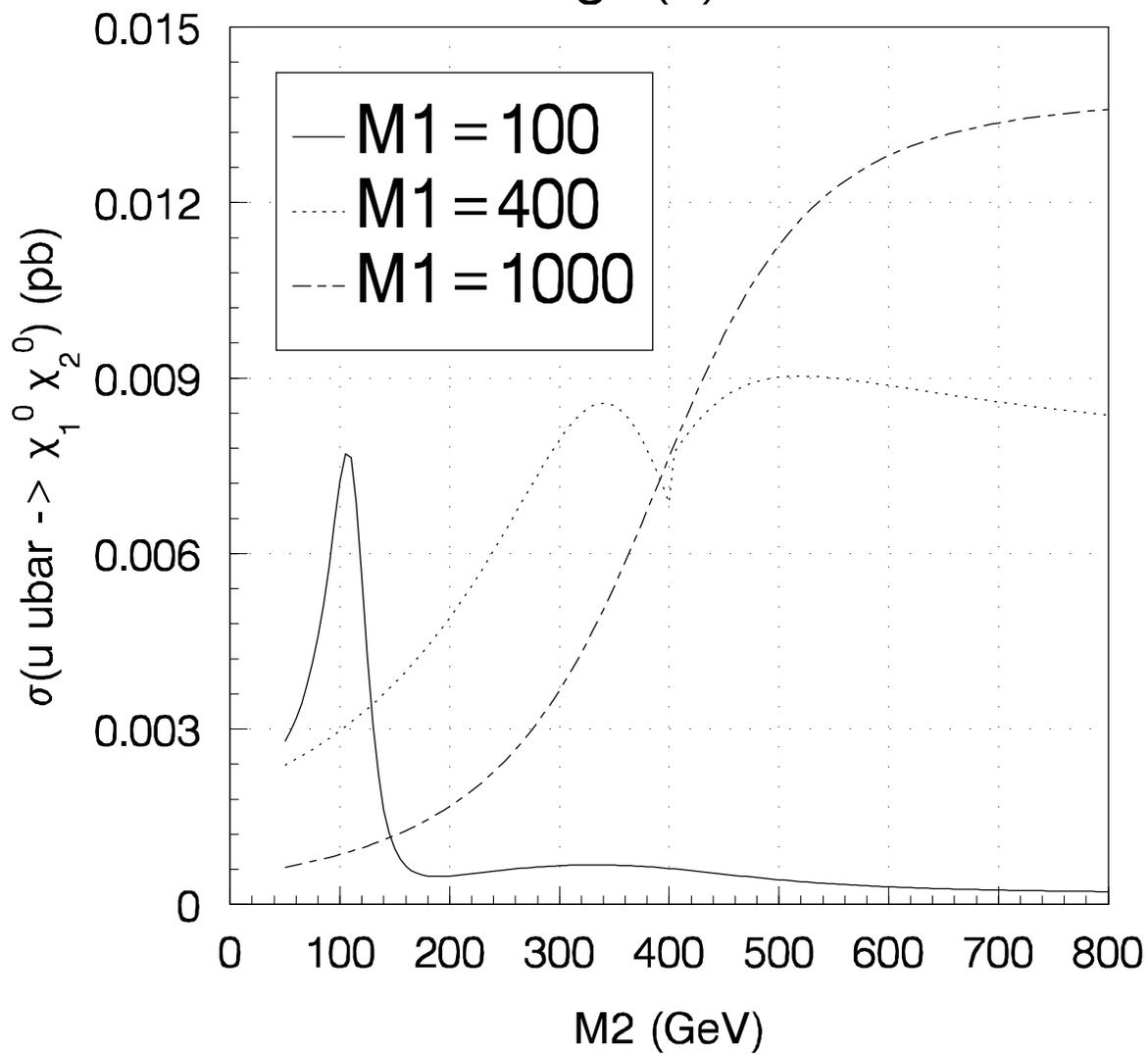
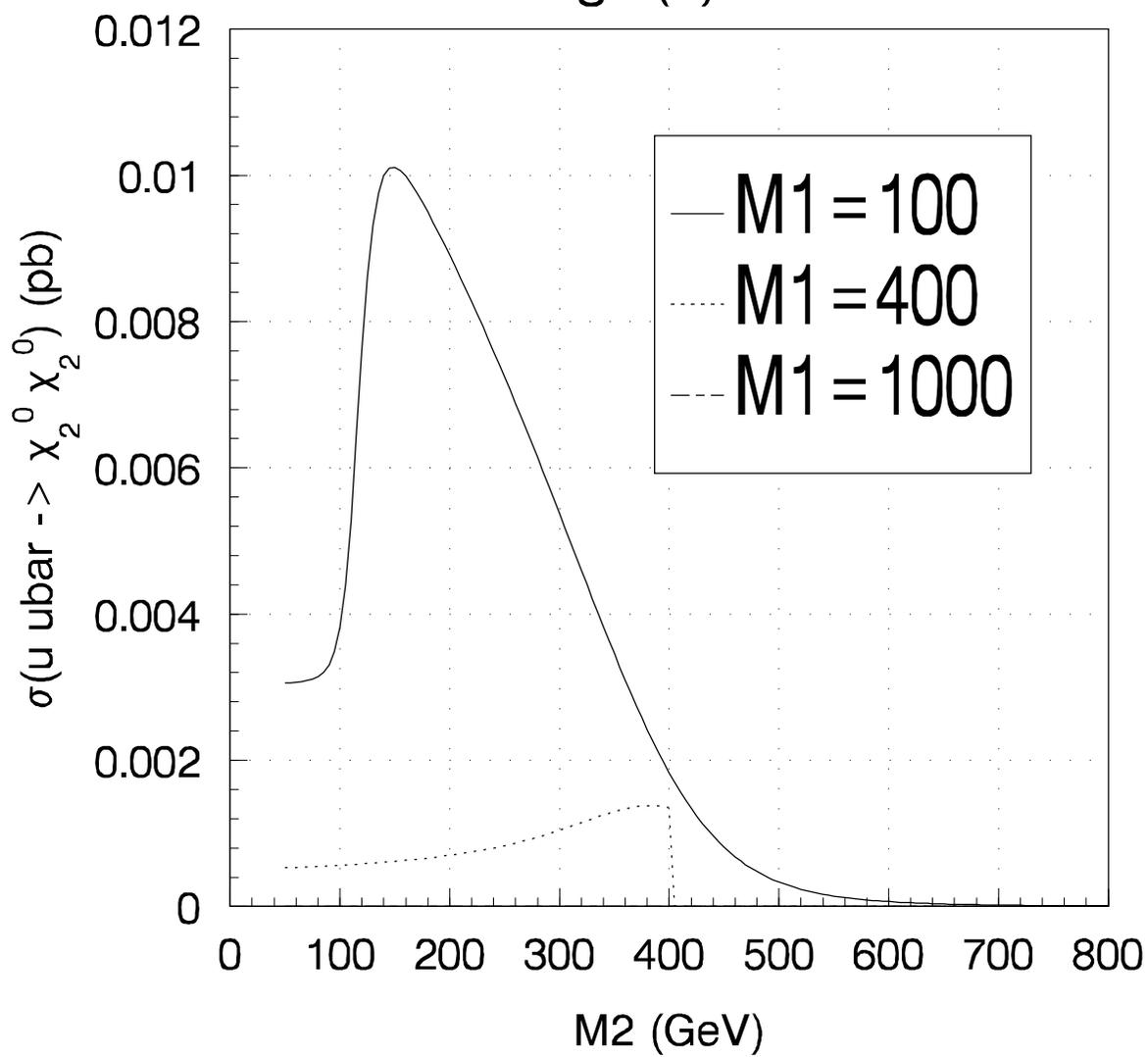
\end{document}